# Enhanced Performance in Epitaxial Graphene FETs with Optimized Channel Morphology

Yu-Ming Lin, Senior Member, IEEE, Damon B. Farmer, Keith A. Jenkins, Senior Member, IEEE, Joseph L. Tedesco, Rachael L. Myers-Ward, Charles R. Eddy, Jr., D. Kurt Gaskill, Yanqing Wu, Member, IEEE, Christos Dimitrakopoulos, Member, IEEE, Phaedon Avouris, Senior Member, IEEE

**Abstract**

This letter reports the impact of surface morphology on the carrier transport and RF performance of graphene FETs formed on epitaxial graphene films synthesized on SiC substrates. Such graphene exhibits long terrace structures with widths between 3-5 μm and steps of 10±2 nm in height. While a carrier mobility above 3000 cm$^2$/Vs at a carrier density of $10^{12}$ cm$^{-2}$ is obtained in a single graphene terrace domain at room temperature, the step edges can result in a vicinal step resistance of ~21 kΩ•μm. By orienting the transistor layout so that the entire channel lies within a single graphene terrace, and reducing the access resistance associated with the ungated part of the channel, a cut-off frequency above 200 GHz is achieved for graphene FETs with channel lengths of 210 nm, which is the highest value reported on epitaxial graphene thus far.

*Index Terms*—graphene, field-effect transistor (FET), radio frequency (RF),

This work is supported by DARPA under contract FA8650-08-C-7838 through the CERA program.

Y.-M. Lin, D.B. Farmer, K. A. Jenkins, Y. Wu, C. Dimitrakopoulos, P. Avouris are with IBM T. J. Watson Research Center, Yorktown Heights, NY 10598 (email: yming@us.ibm.com)

J. L. Tedesco, R. L. Myers-Ward, C. R. Eddy, Jr., D. K. Gaskill are with the Naval Research Laboratories, Washington, DC 20375 USA

The views, opinions, and/or findings contained in the article are those of the authors and should not be interpreted as representing the official views of the DARPA or the Department of Defense.

Graphene, a monolayer of graphite, has received significant attention as a promising candidate for future high-speed electronics because of its high carrier mobility and saturation velocity [1-3]. Since the first demonstration of graphene transistors for radio-frequency (RF) applications [4,5], the performance of graphene RF transistors has been rapidly improved, with cut-off frequencies greater than 100 GHz recently reported by several groups [6-8]. Novel RF functions based on its ambipolar transport properties have also been extensively explored as high-frequency mixers and multipliers[9-11]. The successful development of graphene-based electronics relies on the availability of large-area and high-quality materials. Two types of graphene, epitaxial[12-14] and chemical-vapor-deposited (CVD)[15], have been proposed for wafer-scale production. Epitaxial graphene is produced by thermal decomposition of silicon carbide (SiC) to form uniform graphene layers on the surface of SiC. CVD graphene can also be grown on large area metal surfaces (e.g. Ni, Ru, and Cu), and can be transferred to a substrate of choice for device fabrication. In spite of the exceptional electrical properties of graphene, the performance of graphene devices is strongly affected by both the morphology of the film and its interaction with the environment (e.g. substrate and dielectric interfaces). The performance of CVD graphene devices is particularly sensitive to the transfer process, which can result in the presence of wrinkles, lattice defects, and chemical impurities. Epitaxial graphene typically yields larger domain sizes and does not require transfer to another substrate. However, the morphology of as-prepared epitaxial graphene usually contains steps [14] and other morphological features such as domain boundaries, which may have a negative impact on device performance.

The impact of this morphology on carrier transport in epitaxial graphene is studied here, and it is found that a single terrace step with height of ~ 10 nm can lead to a resistance as high as 21 k$\Omega$•μm. To avoid performance degradation due to such steps, a transistor layout is designed so that the entire device channel lies within a single terrace or domain. Combined with a reduction of access resistances associated with the ungated channel regions, cut-off frequencies above 200 GHz are achieved, demonstrating improved RF performance compared to graphene FETs without such channel morphology and gate structure optimization.

Epitaxial graphene is synthesized on 2-inch semi-insulating 6H-SiC (II-VI, Inc.) wafers. Prior to the graphene growth, the wafer is etched in hydrogen at 1620°C to remove ~ 200 nm of SiC. Graphene is then grown on the Si-face of the substrate at 1620°C in a flowing Ar ambient for 150 minutes at 50 mbar, yielding films consisting of one to two layers of graphene[16]. The morphology of the as-prepared graphene/SiC surface is characterized by atomic force microscopy (AFM) (see Fig. 1(a)), revealing wide and smooth terraces that extend long distances (> 50 μm) parallel to the step edges. The width of these terraces range between 3 and 5 μm, and the step height is 10 nm ±2 nm. This step height is much larger than the substrate steps (~ 1 nm) due to finite miscut angles, and is the result of SiC step bunching due to the thermal anneal[14]. However, the graphene films form continuous over-layers across the step edges, as previously determined by transmission electron microscopy (TEM) studies[13].

To examine the impact of the steps on carrier transport in graphene, two types of two-terminal graphene devices with orthogonal channel orientations are fabricated, as illustrated by the regions outlined and labeled as I and II in Fig. 1(a). The channels of resistor I are parallel to the step edge and located within a single terrace, while those of resistor II contain one step edge along the transport direction. Alignment of the channels with respect to the steps is achieved by depositing Au alignment marks (1 μm x 1μm) and using AFM imaging to map out the surface morphology relative to these marks. In both types of devices, the channel dimensions are all 500 nm in width and 2 μm in length. Subsequent resistance measurements are performed at room temperature on more than 10 devices, yielding distinct resistance values for both device types. Device I possesses an average resistance of 5 kΩ, significantly lower than the average resistance of 48 kΩ for Device II. This clearly shows that carrier transport is severely affected by scattering associated with terrace steps within the channel, which leads to an additional channel resistance as high as 21 kΩ•μm for a single step.

To eliminate the above non-ideal factors associated with the steps, top-gated Hall bar devices are fabricated on a single terrace domain, as depicted by the channel outline III in Fig. 1(a). Fig. 1(b) shows a scanning electron microscopy (SEM) image of such a device, where the entire channel lies within a single terrace. The gate dielectric used here consists of a 10-nm-thick interfacial polymer layer and 10 nm of $HfO_2$ grown by atomic layer deposition (ALD)[17]. At $V_g$ = 0 V, the graphene film is *n*-type with a carrier concentration of ~ $5.6 \times 10^{12}$ $cm^{-2}$ as determined by Hall measurements. Fig. 1(c) shows the measured carrier mobility ($\mu_e$) as a function of carrier density (*n*), where $\mu_e$ increases from 1500 $cm^2/Vs$ to 3300 $cm^2/Vs$ as *n* decreases from $7.1 \times 10^{12}$ $cm^{-2}$ to $1.3 \times 10^{12}$ $cm^{-2}$ by changing the gate voltage. This dependence of carrier mobility on density suggests that the channel consists of single-layer graphene with carrier transport limited by short-range scattering[18], such as neutral impurities and defects.

To evaluate the RF performance of step-free graphene films, top-gated graphene field-effect transistors (GFETs) are fabricated using the top-down approach described in Ref. [6]. Fig. 2(a) shows a schematic of the device cross-section. To achieve optimal performance, the graphene channels are placed within single terraces. These channels are defined by using PMMA as a protective mask, and graphene outside of the channel region is removed by oxygen plasma. The gate dielectric described above (10 nm polymer/10 nm $HfO_2$) is also used here. Both source/drain contacts and gate electrodes are formed by e-beam deposition of Pd/Au (20 nm/40 nm). In addition, to eliminate access resistance due to the ungated channel regions, the gate electrode is designed to overlap with the source/drain electrodes by ~ 50 nm. Fig. 2(b) shows an SEM image of a dual-channel GFET with a channel length of 750 nm, where both channels lie within a single terrace.

Figures 3(a) and (b) show the output characteristics of GFETs with channel lengths ($L_C$) of 750 nm and 210 nm, respectively, measured at room temperature under ambient conditions. The gate voltage is swept between -4 V and 4 V with a step of 1 V. The current density increases with decreasing $L_C$, as expected. For a channel length of 210 nm, the current density is more than 2 mA/μm at $V_d$ = 1.6 V and $V_g$ = 4 V (Fig. 3(b)). The current modulation ratio, defined by $I_d(V_g = 4V)/I_d(V_g = -4V)$, is ~ 3 for channel lengths of 750 nm, and decreases to ~ 2 for $L_C$ = 210 nm. The reduction of the current modulation as $L_C$ decreases is due to the contact resistance between graphene and metal electrodes, which becomes more dominant as $L_C$ decreases. From Fig. 3(b), the total contact resistance (2$R_C$) is estimated to be ~ 700 Ω•μm. Figs. 3(c) and (d) show the transfer characteristics at $V_d$ = 1.6 V for $L_C$ = 750 nm and 210 nm, respectively. The corresponding transconductances ($g_m$) of these devices are plotted on the right axes of Figs. 3(c)(d). These transconductances exhibit maximum values at $V_g$ ~ 0 V of 210 μS/μm and 320 μS/μm for $L_C$ = 750 and 210 nm, respectively. In comparison, the device reported in Ref. [6], which was fabricated on epitaxial graphene without morphology or structure optimization, possessed a lower current density of ~ 1 mA/μm and a maximum $g_m$ of only ~ 200 μS/μm at $V_d$ = 1.6 V for $L_C$ = 240 nm.

The RF performance of the GFETs are characterized by on-chip S-parameter measurements up to 30 GHz. Standard de-embedding procedures using specific "short" and "open" structures with identical layouts are employed to remove the effects of parasitic capacitances and the resistances associated with the pads and connections. Figure 4 plots the measured current gain |h$_{21}$| as a function of frequency for GFETs with different $L_C$ at $V_d$ = 2.5 V and $V_g$ = 0 V after de-embedding. The current gain |h$_{21}$| follows the expected 1/$f$ frequency dependence, and the cut-off frequency ($f_T$) is determined by the extrapolation of |h21|, yielding $f_T$ of 60 GHz and 210 GHz for GFETs with $L_C$ = 750 nm and 210 nm, respectively. Because of the de-embedding, this cutoff frequency represents the response of the intrinsic device under the gate electrode. The $f_T$ of 210 GHz is the highest value reported for epitaxial GFETs thus far. Previously, $f_T$ = 100 GHz was achieved for GFETs with $L_c$ = 240 nm using a similar gate dielectric[6]. This illustrates the performance enhancement afforded by the optimization of channel morphology and device structure presented in this study. The frequency-length product of these graphene FETs are ~ 45 GHz•um, which is comparable to the highest value of ~ 43 GHz•μm obtained from exfoliated graphene [7] and much higher than those of Si (~ 8.9 GHz•μm) and InP (~ 17GHz•μm) [19].

In conclusion, the impact of surface morphology on carrier transport and RF performance of epitaxial graphene films synthesized on SiC substrates is reported. Carrier mobility above 3000 cm$^2$/Vs at a carrier density of 10$^{12}$ cm$^{-2}$ is obtained on a single graphene terrace at room temperature. Step edges, however, contribute an additional interfacial resistance of ~21 kΩ•μm for a single step along the transport direction, and also degrades RF performance. By forming the transistor so that its entire channel lies within a single graphene terrace, and reducing the access resistance associated with ungated channel regions, a cut-off frequency above 200 GHz is achieved for a channel length of 210 nm.

Steps and other topographical features such as domain boundaries and surface wrinkles will influence graphene transport not only on SiC epitaxial graphene, but even in CVD grown and transferred graphene. It is therefore imperative that the effect of these features is understood and the morphology of the graphene layer is controlled.

## Acknowledgment

The authors would like to thank A. Grill, A. Valdes-Garcia, C. Y. Sung, F. Xia, and W. Zhu for insightful discussions, and B. Ek and J. Bucchignano for technical assistance. We thank DARPA for financial support through the CERA program (contract FA8650-08-C-7838).

# REFERENCES

[1] K. Bolotin, K. Sikes, Z. Jiang, M. Klima, G. Fudenberg, J. Hone, P. Kim, and H. Stormer, "Ultrahigh electron mobility in suspended graphene," *Solid State Communications*, vol. 146, 2008, pp. 351-355.

[2] A. K. Geim and K.S. Novoselov, "The rise of graphene.," *Nature materials*, vol. 6, 2007, pp. 183-91.

[3] P. Avouris, "Graphene: Electronic and Photonic Properties and Devices," *Nano Letters*, vol. 10, 2010, p. 4285.

[4] Y.-M. Lin, K.A. Jenkins, A. Valdes-garcia, J.P. Small, D.B. Farmer, and P. Avouris, "Operation of Graphene Transistors at Gigahertz Frequencies 2009," *Nano letters*, vol. 9, 2009, pp. 422-426.

[5] I. Meric, N. Baklitskaya, P. Kim, and K.L. Shepard, "transistors," *IEEE InternationalElectron Devices Meeting, 2008. IEDM Technical Digest*, 2008, pp. 1-4.

[6] Y.-M. Lin, C. Dimitrakopoulos, K. A. Jenkins, D.B. Farmer, H.-Y. Chiu, A. Grill, and P. Avouris, "100-GHz transistors from wafer-scale epitaxial graphene," *Science*, vol. 327, 2010, p. 662.

[7] L. Liao, Y.-C. Lin, M. Bao, R. Cheng, J. Bai, Y. Liu, Y. Qu, K.L. Wang, Y. Huang, and X. Duan, "High-speed graphene transistors with a self-aligned nanowire gate," *Nature*, vol. 467, 2010, pp. 305-308.

[8] Y. Wu, Y.-M. Lin, A. a Bol, K. a Jenkins, F. Xia, D.B. Farmer, Y. Zhu, and P. Avouris, "High-frequency, scaled graphene transistors on diamond-like carbon," *Nature*, vol. 472, 2011, pp. 74-78.

[9] H. Wang, S. Member, A. Hsu, J. Wu, J. Kong, and T. Palacios, "Graphene-Based Ambipolar RF Mixers," *IEEE Electron Device Letters*, vol. 31, 2010, pp. 906-908.

[10] J.S. Moon, D. Curtis, D. Zehnder, S. Kim, D.K. Gaskill, G.G. Jernigan, C.R. Eddy, P.M. Campbell, K. Lee, and P. Asbeck, "Low-Phase-Noise Graphene FETs in Ambipolar RF Applications," *IEEE Electron Device Letters*, vol. 32, 2011, pp. 270-272.

[11] X. Yang, G. Liu, A. a Balandin, and K. Mohanram, "Triple-mode single-transistor graphene amplifier and its applications," *ACS nano*, vol. 4, 2010, pp. 5532-8.

[12] C. Berger, Z. Song, X. Li, X. Wu, N. Brown, C. Naud, D. Mayou, T. Li, J. Hass, A.N. Marchenkov, E.H. Conrad, P.N. First, and W. A. de Heer, "Electronic confinement and coherence in patterned epitaxial graphene.," *Science*, vol. 312, 2006, pp. 1191-6.

[13] C. Dimitrakopoulos, Y.-M. Lin, A. Grill, D.B. Farmer, M. Freitag, Y. Sun, S.-J. Han, Z. Chen, K. a Jenkins, Y. Zhu, Z. Liu, T.J. McArdle, J. A. Ott, R. Wisnieff, and P. Avouris, "Wafer-scale epitaxial graphene growth on the Si-face of hexagonal SiC (0001) for high frequency transistors," *Journal of Vacuum Science & Technology B: Microelectronics and Nanometer Structures*, vol. 28, 2010, p. 985.

[14] K.V. Emtsev, A. Bostwick, K. Horn, J. Jobst, G.L. Kellogg, L. Ley, J.L. McChesney, T. Ohta, S. a Reshanov, J. Röhrl, E. Rotenberg, A.K. Schmid, D. Waldmann, H.B. Weber, and T. Seyller, "Towards wafer-size graphene layers by atmospheric pressure graphitization of silicon carbide.," *Nature materials*, vol. 8, 2009, pp. 203-7.

[15] X. Li, W. Cai, J. An, S. Kim, J. Nah, D. Yang, R. Piner, A. Velamakanni, I. Jung, E. Tutuc, S.K. Banerjee, L. Colombo, and R.S. Ruoff, "Large-area synthesis of high-quality and uniform graphene films on copper foils.," *Science*, vol. 324, 2009, pp. 1312-4.

[16] D.K. Gaskill, G. Jernigan, P. Campbell, J.L. Tedesco, J. Culbertson, B. VanMil, R.L. Myers-Ward, C. Eddy Jr., J. Moon, D. Curtis, M. Hu, D. Wong, C. McGuire, J. Robinson, M. Fanton, T. Stitt, D. Snyder, X. Wang, and E. Frantz, "Epitaxial Graphene Growth on SiC Wafers," *ECS Transactions*, vol. 19, 2009, pp. 117-124.

[17] D.B. Farmer, H.-Y. Chiu, Y.-M. Lin, K. a Jenkins, F. Xia, and P. Avouris, "Utilization of a buffered dielectric to achieve high field-effect carrier mobility in graphene transistors.," *Nano letters*, vol. 9, 2009, pp. 4474-8.

[18] W. Zhu, V. Perebeinos, M. Freitag, and P. Avouris, "Carrier scattering, mobilities, and electrostatic potential in monolayer, bilayer, and trilayer graphene," *Physical Review B*, vol. 80, 2009, p. 235402.

[19] ITRS roadmap. http://www.itrs.net/Links/2008ITRS/Home2008.htm

Figures

Fig. 1 (a) AFM image of graphene grown on the Si-face of SiC substrate, showing the terrace structure and steps. The scanning area is 21μm x 21μm. The solid lines sketch the outline of the graphene channels of different devices used to study the impact of terrace steps on carrier transport. Channels I and III both lie within a single terrace while the channel II contains one step along the transport direction. (b) SEM image of a gated Hall bar structure fabricated on a single terrace. The graphene channel is highlighted by the dashed line. (c) Carrier mobility measured as a function of carrier density for the device shown in (b).

Fig. 2 (a) Schematic cross-section of a top-gated GFET. To reduce the access resistance associated with the ungated regions, the gate electrodes overlap with the source and drain by ~ 50 nm, which is mainly limited by the resolution and accuracy of the e-beam lithography. (b) (false color) SEM image of a dual-channel graphene FET where both channels are located within a single terrace. The channel length is 750 nm and the total channel width is 30 μm.

Fig. 3 (a)(b) Measured output characteristics of GFETs with different channel lengths: (a) 750 nm and (b) 210 nm. (c)(d) Measured transfer characteristics of these graphene FETs at $V_d$ = 1.6 V. The transconductance is shown on the right axes as a function of gate voltage.

Fig. 4 Measured current gain $|h_{21}|$ of GFETs at $V_d$ = 2.5 V, showing the 1/$f$ frequency dependence

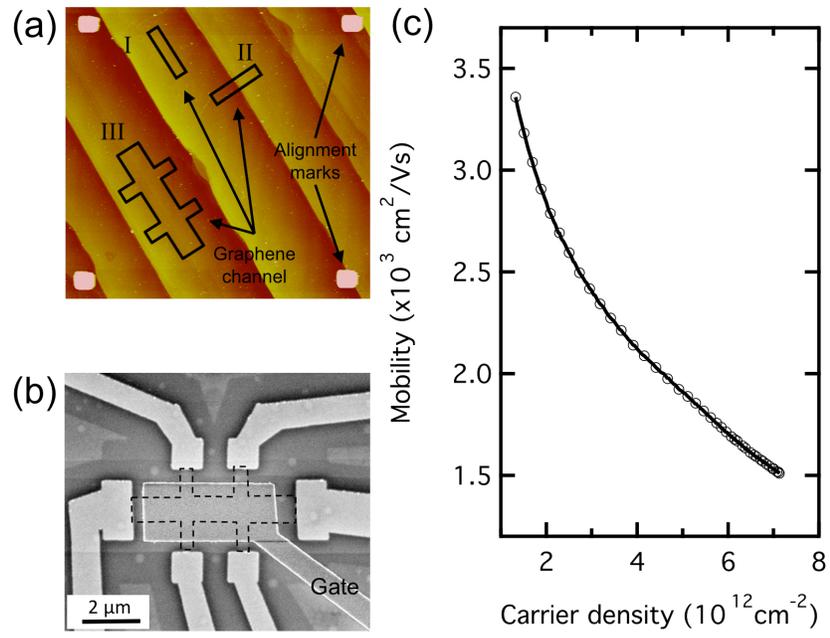

Fig. 1

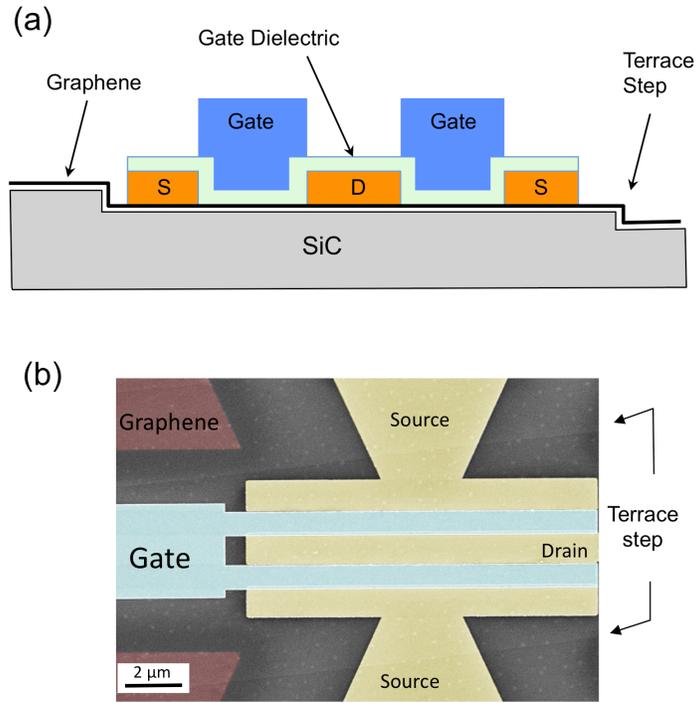

Fig. 2

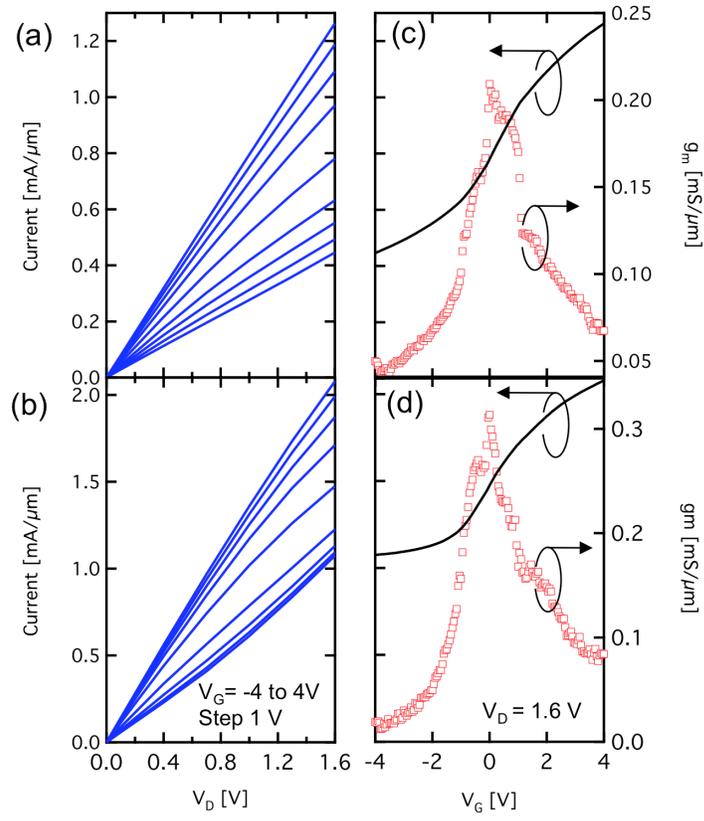

Fig. 3

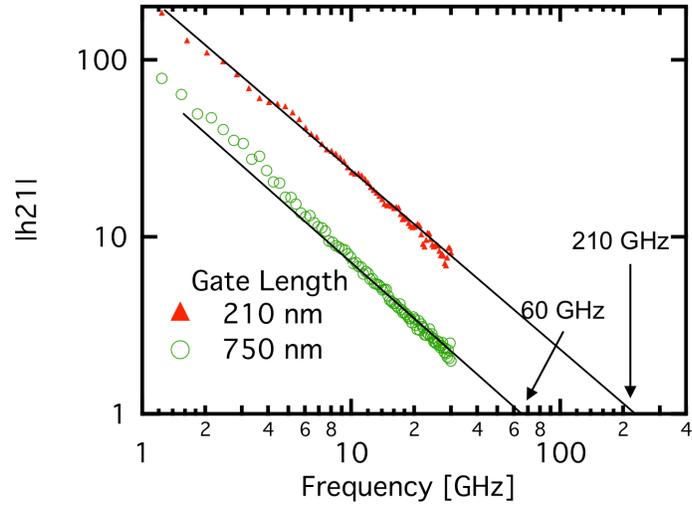

Fig. 4